\documentclass{aa} 

\def\lapp{\ifmmode\stackrel{<}{_{\sim}}\else$\stackrel{<}{_{\sim}}$\fi}
\def\gapp{\ifmmode\stackrel{>}{_{\sim}}\else$\stackrel{>}{_{\sim}}$\fi}
\newcommand{\EPSRA}{J1842--0415}
\newcommand{\EPSRB}{J1844--0310}
\newcommand{\EPSRC}{J1845--0316}
\newcommand{\EPSRD}{J1841--0345}

\begin{document}

\title{A 1400-MHz pilot search for young pulsars}

\author{D.R.~Lorimer\inst{1,2}, M.~Kramer\inst{1,3}, P.~M\"uller\inst{1}, 
        N.~Wex\inst{1}, A.~Jessner\inst{1}, C.~Lange\inst{1},
        R.~Wielebinski\inst{1}}
 
\offprints{D.R.~Lorimer (dunc@naic.edu)}
\institute{Max-Planck-Institut f\"ur Radioastronomie, Auf dem H\"ugel 69,
             D-53121 Bonn, Germany 
 \and       Present address: National Astronomy and Ionospheric Center, 
	    Arecibo Observatory, HC3 Box 53995, PR 00612, USA
 \and       Present address: University of Manchester, Jodrell Bank 
            Observatory, Macclesfield, Cheshire, SK11 9DL, UK
}

\thesaurus{03.20.9 06(03.20.3 08.16.7 \EPSRA \, 08.16.7 
\EPSRB \, 08.16.7 \EPSRC \, 08.16.7 \EPSRD)}
\authorrunning{Lorimer et al.}
\titlerunning{A 1400-MHz pulsar search}
\date{\today}
\maketitle

\begin{abstract} 
We have used the Effelsberg 100-m radio telescope to conduct a
1400-MHz ($\lambda$ 21-cm) search for young and rapidly rotating radio
pulsars along a 2 deg$^2$ strip of the northern Galactic plane defined
by $28^{\circ} \leq l \leq 30^{\circ}$ and $|b|\leq0.5^{\circ}$. This
region lies close to the Scutum spiral arm which is already known to
contain a number of radio and X--ray pulsars. The search was nominally
sensitive to pulsars with 1400-MHz flux densities above 0.3 mJy; this
represents a threefold improvement in sensitivity over all previous
searches of this region of the Galaxy.  Four new long-period pulsars
were discovered as a result of this survey.  All three previously
known pulsars in this region were also detected. The four new pulsars
are relatively young ($<$ 0.5 Myr), weak ($<1$ mJy) sources with
dispersion measures in the range 170--910 cm$^{-3}$ pc.  None of the
newly-discovered pulsars are associated with catalogued supernova
remnants.

\end{abstract}

\section{Introduction}
\label{sec:intro}

Although the majority of pulsar radio flux density spectra peak at
frequencies around 400 MHz, during the 1980s it was realised that the
sensitivity of pulsar surveys conducted at such frequencies becomes
seriously compromised when searching along the Galactic plane.  The
reasons for this are twofold: (1) The system temperature becomes
dominated by the sky background radiation. Typical 400-MHz sky
background temperatures are $\sim900$ K in the direction of the
Galactic centre, and $\sim300$ K along the Galactic plane
\nocite{hssw82} (Haslam et al.~1982). (2) The observed pulse width can
become much larger than the intrinsic width due to multi-path
scattering and/or dispersion by free electrons in the interstellar
medium. Both these effects lead to a net reduction in signal-to-noise
ratio. In extreme cases of scattering and dispersion, the observed
pulse width becomes comparable to the pulse period and the pulsar is
no longer visible as a periodic radio source.

Fortunately, all these effects diminish strongly at a higher observing
frequency: The brightness temperature of the radio continuum emission
$T_{\nu}$ at a given observing frequency $\nu$ has a power law
dependence $T_{\nu} \propto \nu^{\beta}$ with a spectral index
$\beta\sim-3$ (Lawson et al.~1987; Reich \& Reich 1988).
\nocite{lmop87,rr88} This means that the 408-MHz sky background
temperatures quoted above are reduced by more than an order of
magnitude for high frequency ($\gapp 1$-GHz) surveys. Furthermore,
pulse dispersion and scattering scale as $\Delta \nu/\nu^{3}$ and
$\nu^{-4}$ respectively (e.g.~Manchester \& Taylor 1977),
\nocite{mt77} for an observing frequency $\nu$ and bandwidth $\Delta
\nu$.

Clifton \& Lyne (1986) (see also Clifton et al.~1992)
\nocite{cl86,clj+92} were the first to really demonstrate the worth of
surveying at high frequencies. In a 1400-MHz survey of a thin strip of
200 deg$^2$ along the Galactic plane, Clifton et al.~found 40 new
pulsars. All of these sources were missed by a previous 390-MHz survey
(Stokes et al.~1985) \nocite{stwd85} which overlapped the same
region. This was in spite of the fact that, after scaling the
sensitivity limits for typical pulsar spectral indices, the Stokes et
al.~survey had twice the nominal sensitivity of the Clifton et
al.~survey.  Johnston et al.~(1992a) carried out a complementary
survey of the southern Galactic plane using the Parkes radio telescope
at 1520 MHz, \nocite{jlm+92} finding 46 pulsars missed by previous
lower frequency searches covering this region \nocite{mlt+78}
(Manchester et al.~1978).

The pulsars discovered in these two high frequency surveys are
primarily young neutron stars that have not had time to move far from
their birth places close to the Galactic plane.  A large sample of
such objects is desirable for studies of the birth and evolution of
neutron stars and of the size of the neutron star population in the
inner Galaxy \nocite{joh94} (Johnston 1994). In addition, these
surveys discovered several interesting binary pulsars including PSR
B1259$-$63 --- a 48-ms pulsar in a 3.4-yr orbit around a 10
M$_{\odot}$ Be star \nocite{jml+92} (Johnston et al.~1992b).

Significant improvements in sensitivity have lead to renewed interest
in Galactic plane searches. In particular Camilo et al.~(2000a)
\nocite{clm+00} report the discovery of over 400 pulsars in the first
half of a new survey of the southern Galactic plane using the recently
commissioned Parkes $\lambda$ 21-cm multibeam system. Their survey is
some seven times more sensitive than the Clifton et al.~and Johnston
et al.~surveys, and the new discoveries already include several binary
pulsars (e.g.~Lyne et al.~1999), as well as a large number of very
distant, high dispersion measure, sources.

A preliminary account of the exciting results from the Parkes multibeam 
survey (Camilo et al.~1997) prompted us to utilise the large collecting
area of the 100--m Effelsberg radio telescope to perform a new search
along the northern Galactic plane.  In this paper we report on a small
survey carried out during 1998 to test the feasibility of future
observations with a wide-bandwidth search system currently under
development. This pilot search proved successful, discovering four new
pulsars --- the first ever found with this telescope.  In
Sect.~\ref{sec:obs} we describe in some detail the survey observations
and data reduction techniques. In Sect.~\ref{sec:sens} we estimate the
sensitivity of the survey. The results are presented in
Sect.~\ref{sec:res}. These results, along with follow-up timing
observations, are discussed in Sec.~\ref{sec:fup}. Finally, in
Sect.~\ref{sec:conc}, we summarise the main conclusions from this work
and their implications for future pulsar search experiments at Effelsberg.

\section{Survey observations and data reduction}
\label{sec:obs}

All observations reported in this paper were carried out at a centre
frequency of 1402 MHz on a number of separate sessions between 1998
June and 1999 April using the 100--m Effelsberg radio telescope
operated by the Max-Planck-Institut f\"ur Radioastronomie.  Although
1400-MHz timing observations at Effelsberg are routinely made with
typical bandwidths of 40 MHz or more \nocite{kxl+98} (see e.g.~Kramer
et al.~1998), the search hardware available to us has a maximum bandwidth of
16 MHz in each of the two orthogonal, circular polarisation channels.
Nonetheless, the large forward gain of the telescope at 1400 MHz (1.5
K Jy$^{-1}$), the relatively low system temperature of the receiver
(35 K) and long integration times employed in the survey (35 min per
pointing) means that the system achieves a sensitivity which
represents a threefold improvement over that achieved by Clifton et
al.~(1992) during their survey.

The main aim of the observations reported here was to test the
feasibility of a larger search with a 100-MHz bandwidth system which
is presently being commissioned. Given the limited amount of telescope
time available for this pilot project, we chose to restrict our search
area to a 2 deg$^2$ patch of the Galactic plane defined by
$28^{\circ}<l<30^{\circ}$ and $|b|<0.5^{\circ}$. The rationale for
this choice is simple --- this line-of-sight is close to the Scutum
spiral arm and as a result passes through one of the most pulsar-rich
parts of the northern Galactic plane. In addition, since this part of
the sky is not visible from Arecibo, Effelsberg is presently the
largest radio telescope in the world capable of surveying it.  The
survey region was divided up into a grid of 126 positions consisting
of 9 strips of 14 positions along lines of constant galactic latitude
($b=0.0^{\circ}, \pm0.12^{\circ}, \pm0.24^{\circ}, \pm0.36^{\circ},
\pm0.48^{\circ}$).  This choice of spacing ensured some overlap
between the 3-dB width of the telescope beam (9').  The
$b=0.0^{\circ}$ strip was centred on $l=29^{\circ}$. Beam centres on
adjacent strips were alternately offset by half a beam width to ensure
the most efficient coverage on the sky.

At the start of each observing run, we carried out a 5-min observation
of PSR B2011+38. This relatively luminous 230-ms pulsar has a
dispersion measure of 239 cm$^{-3}$ pc and is known not to be prone to
significant intensity variations due to interstellar scintillation
(Lorimer et al.~1995).  The fact that the search code detected this
pulsar with consistently high signal-to-noise ratios (consistent with
its 1400-MHz flux density --- $6.4\pm0.5$ mJy; Lorimer et al.~1995) gave us
confidence that the individual filterbank channels were functioning
normally, and that the nominal system sensitivity was being achieved.

The search field is visible from Effelsberg for about 8.5 hr per
day. Since each grid position in the field was observed for 35 min,
we typically observed up to 14 separate positions on the sky
during a given transit. In search mode, the incoming signals of each
polarisation are fed into a pair of $4\times4$-MHz filterbanks. The
outputs from the filterbanks are subsequently detected and digitised
every 500 $\mu$s using 2-MHz voltage-to-frequency converters,
resulting in an effective 10-bit quantisation of the signals. 
This is the fastest sustainable data rate using this system. Signals
from the orthogonal polarisations were combined to form a total power
time series for each 4-MHz frequency channel over the band.  These
four frequency channels are then passed to the standard Effelsberg
Pulsar Observing System (see \nocite{kra95} Kramer 1995) which stored 
contiguous blocks of data to disk every 1024 samples (0.512 s).

The four-channel search system used for this survey results in
refreshingly low data rates compared to most other searches where the
backends routinely sample 256 channels or more \nocite{mld+96} (see
e.g.~Manchester et al.~1996). The main advantage of such a simple
system is that, as soon as each 35-min integration was complete, a
preliminary analysis of the data could be carried out well within the
time that the telescope was observing the next grid position. This
quasi on-the-fly processing scheme allowed rapid re-observation of any
pulsar candidates found during the search.

The data analysis procedure was optimised to search Fourier spectra of
each 35-min time series for dispersed periodic signals. The software
for this purpose was developed largely from scratch, taking advantage
of ideas used to process our on-going search of the Galactic centre
from Effelsberg (Kramer et al.~1996; Kramer et al.~2000), as well as
previous experience gained by one of us (DRL) during the Parkes
$\lambda$ 70-cm Southern Sky Survey (Manchester et al.~1996; Lyne et
al.~1998). \nocite{mld+96,lml+98} We also made use of several
``standard'' pulsar search techniques described in detail by a number
of authors \nocite{hr75,lyn88,nic92} (Hankins \& Rickett 1975; Lyne
1988; Nice 1992). In what follows we give a brief overview of our
analysis procedure.

\begin{figure*}[hbt]
\setlength{\unitlength}{1in}
\begin{picture}(0,3.7)
\put(0.5,4.4){\includegraphics{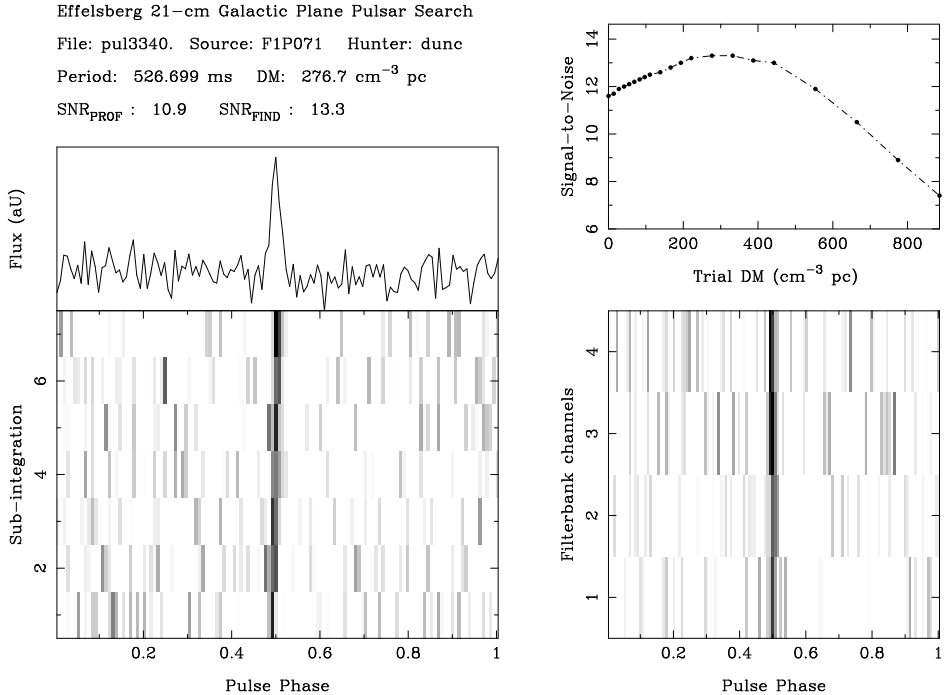}}
\end{picture}
\caption[]{
Sample search code output showing the discovery observation of PSR
\EPSRA --- the first of the four pulsars found during the
survey. This plot shows how a typical pulsar appears to the search
code and summarises the various diagnostics we used to 
identify the best pulsar candidates from the survey (see text).
}
\label{fig:epsr1} 
\end{figure*}

We adopted a two-stage data analysis procedure whereby the data were
quickly analysed after the observation at Effelsberg and then stored
on magnetic tape for more detailed off-line analyses at Bonn. The three
differences in the analyses are the range of dispersion measures
searched, the signal-to-noise thresholds, and the method used to excise
radio frequency interference (see below).

Both analyses begin by computing the Fast Fourier Transform
of the $2^{22}$ point time series in each of the four frequency
channels.  Since the dispersion measure (DM) of any pulsar is a priori
unknown we need to de-disperse the data for a number of
trial DM values before the periodicity search begins. For our purposes
this is most readily achieved by applying the shift theorem
\nocite{bra65} (see e.g.~Bracewell 1965) to the Fourier components of each
channel before summing appropriately to produce a number of
de-dispersed amplitude spectra. Our on-line analysis in Effelsberg
produced 18 amplitude spectra for each beam corresponding to a DM range
between zero and 1,500 cm$^{-3}$ pc. Subsequent analyses in
Bonn produced, in addition to this, a further 18 spectra per beam
which increased the range of DMs out to 10,000 cm$^{-3}$ pc. Each
amplitude spectrum was then searched for harmonically related spikes in
the Fourier domain --- the characteristic signature of any periodic
signal. Since pulsar signals have generally short duty cycles, and
therefore many harmonics, we summed the spectra over 2, 4, 8, and 16 
harmonics using an algorithm described by Lyne (1988) \nocite{lyn88} 
and repeated the search for significant spectral features.

Having completed the search of all the amplitude spectra for a
given beam, we then compiled a list of all non-harmonically-related
spectral features with a signal-to-noise ratio greater than 8 in the
Effelsberg analyses and 7 in the Bonn analyses. Typically, depending
on the amount of interference present in the data, there are of order
five to ten such ``pulsar candidates'' in each beam. For each
candidate, the analysis described so far resulted in a period $P$ and
dispersion measure DM; the latter quantity is based upon the maximum
spectral signal-to-noise ratio found as a function of all the DM
trials.  Working now in the time domain, we fold the filterbank
channels at the nominal period of each candidate to produce one pulse
profile per channel. These profiles are then de-dispersed at the
nominal DM to produce an integrated profile over the 16-MHz band.

The results of this analysis are summarised in the plot of the form
shown in Fig.~\ref{fig:epsr1} which is the output from the discovery
observation of PSR \EPSRA.  This plot serves as a good example showing
the characteristics of a strong pulsar candidate. The high
signal-to-noise integrated profile (top left panel) can be seen as a
function of time and radio frequency in the grey scales (lower left
and right panels).  In addition, the dispersed nature of the signal is
immediately evident in the upper right hand panel which shows the
signal-to-noise ratio as a function of trial DM. This combination of
diagnostics proved extremely useful in differentiating between a good
pulsar candidate and spurious interference.

The most significant difference between our two data reduction
strategies concerns the methods employed to eliminate radio frequency
interference. Since the radio frequency environment in Effelsberg is
pervaded by a number of man-made signals with fluctuation frequencies
predominantly between 10--2000-Hz, both modes of data reduction
required some means of excising these unwanted signals. The ``on-line''
data reduction mode in Effelsberg achieved this by simply
clipping all spectral features above 10-Hz whose amplitudes exceeded
five times the spectral rms!  Whilst this simple-minded approach
was sufficient to detect and confirm all the pulsars finally
discovered in the survey, we were aware that it significantly
compromised our sensitivity to pulsars with periods below 0.1 s. 

To address this important issue, our data analysis procedure in Bonn
made use of the fact that the vast majority of man-made interfering
signals are not dispersed and occur predominantly at a constant
fluctuation frequency at any given epoch.  These signals are
immediately apparent in a compilation of a large number of zero-DM
amplitude spectra for different beam positions.  Based on the
statistics of over 60 individual spectra, we constructed a ``spectral
mask'' which contains the frequencies of those spectral features which
occur more than 5 times above a signal-to-noise threshold of 7.  We
found 611 such frequencies between 30 and 2000-Hz --- 0.06\% of the
total number of spectral bins. Most of these are in fact related to
the 50-Hz mains power line.  By masking (i.e.~ignoring) just these
frequencies in our analysis, it was then possible to detect
short-period pulsars with fundamental frequencies outside the masked
frequency bins in our data. We verified the validity of this approach
by a analysing number of test observations on millisecond and short-period
pulsars which were essentially undetectable without the use of the
spectral mask, simply because of the dominating effect of the
interfering signals. Thus, although we did not detect any short-period
pulsars in this survey, we are confident that no potentially
detectable pulsars with fundamental frequencies outside the masked
frequency bins were missed because of radio frequency interference.

\section{Search sensitivity}
\label{sec:sens}

To estimate the sensitivity of this survey, we make use of the following
expression which is similar to that derived by Dewey et al.~(1984) 
\nocite{dss+84} to find the minimum detectable flux density a pulsar has 
to have in order to be detectable:
\begin{equation}
\label{equ:smin}
   S_{\rm min} = \frac{\eta \, T_{\rm sys}}{G \sqrt{2 \Delta
        \nu \tau}} \left(\frac{W}{P-W} \right)^{1/2}.
\end{equation}
Here the constant factor $\eta$ takes into account losses in the
hardware and the threshold signal-to-noise ratio above which a
detection is considered significant ($\eta \approx 10$ in our case),
$T_{\rm sys}$ is the system temperature (see below), $G$ is the gain
of the telescope (1.5 K Jy$^{-1}$ for Effelsberg operating at 21-cm),
$\Delta \nu$ is the observing bandwidth (16-MHz for this survey), the
factor of $\sqrt{2}$ indicates that two polarisation channels were
summed, $\tau$ is the integration time per telescope pointing (35 min), 
$P$ is the period of the pulsar and $W$ is the observed width of the pulse.

The system temperature $T_{\rm sys}$ is essentially the sum of the
noise temperature of the receiver $T_{\rm rec}$, the spillover noise
into the beam side-lobes from the ground $T_{\rm spill}$ and the
excess background temperature $T_{\rm sky}$ caused largely by
synchrotron radiating electrons in the Galactic plane itself. From
regular calibration measurements we found $T_{\rm rec}$ to be 35
K. The spillover contribution $T_{\rm spill}$ was estimated to be 5 K
for typical telescope elevations during survey observations.  We
estimate $T_{\rm sky}$ by scaling the 408-MHz all-sky survey of Haslam
et al.~(1982) \nocite{hssw82} to 1400 MHz assuming a spectral index of
--2.7 (Lawson et al.~1987), \nocite{lmop87} finding a typical value in
the direction $l=29$ and $b=0.0$ to be 15 K.  With these values in
Eq.~(\ref{equ:smin}), we find the minimum flux density for
detecting a 0.5 s pulsar with a duty cycle of 4\% to be about 0.3 mJy.

We caution that this sensitivity estimate should be viewed as a ``best
case scenario'', valid for relatively long-period pulsars with low
dispersion measures and narrow pulses observed at the beam centre.
The effects of sampling and dispersion and pulse scattering
significantly degrade the search sensitivity at short periods.
Specifically, the observed pulse width $W$ in Eq.~(\ref{equ:smin}) is
often likely to be greater than the intrinsic width $W_{\rm int}$
emitted at the pulsar because of the scattering and dispersion of
pulses by free electrons in the interstellar medium, and by the
post-detection integration performed in the receiver. The sampled
pulse profile is the convolution of the intrinsic pulse width and
broadening functions due to dispersion, scattering and integration and
is estimated from the following quadrature sum:
\begin{equation}
  W^2 = W_{\rm int}^2 + t_{\rm samp}^2 + t_{\rm DM}^2 + t_{\rm scatt}^2,
\end{equation} 
where $t_{\rm samp}$ is the data sampling interval, $t_{\rm DM}$ is
the dispersion broadening across one filterbank channel and $t_{\rm
scatt}$ is the interstellar scatter broadening. 

\begin{figure}
\setlength{\unitlength}{1in}
\begin{picture}(0,2.5)
\put(-0.3,2.9){\includegraphics{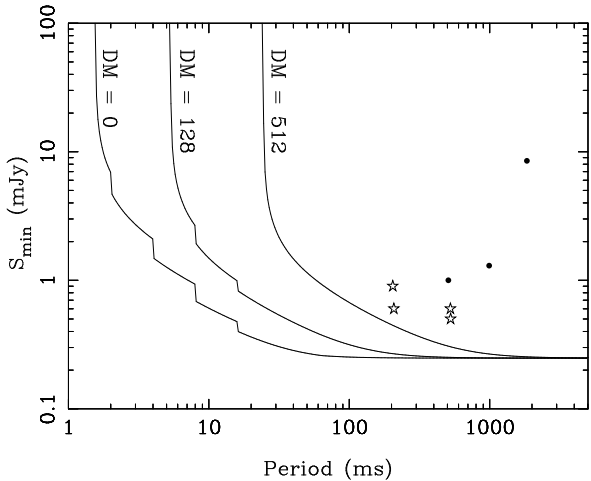}}
\end{picture}
\caption[]
{
The limiting sensitivity of the survey as a function of pulse period.
The scattering times used in these calculations were estimated from
the Taylor \& Cordes (1993) electron density model for each DM for the
Galactic coordinates ($l=29^{\circ}; b=0^{\circ}$). Solid points denote
the three previously known pulsars. Stars show the four 
newly-discovered pulsars.
}
\label{fig:smin}
\end{figure}

To highlight the effects of pulse broadening on sensitivity, in
Fig.~\ref{fig:smin} we present the effective sensitivity as a function
of period for a hypothetical pulsar with an intrinsic duty cycle of
5\% for assumed DMs of 0, 128 and 512 cm$^{-3}$\,pc. The scallops in
the curves at short periods reflect the reduction in sensitivity due to
the loss of higher-order harmonics in the Fourier spectrum (see
e.g.~Nice 1992). \nocite{nic92} The severe degradation in sensitivity
at short periods and high dispersion measures is clearly seen in this
diagram.  In particular, we note that due to the dispersion across
individual filterbank channels, the present observing system is
essentially insensitive to pulsars with periods less than 30 ms and
DMs larger than 500 cm$^{-3}$\,pc. 

In the discussion hitherto we have implicitly assumed that the
apparent pulse period remains constant during the observation. Given
the necessarily long integration times employed to achieve good
sensitivity, this assumption is only valid for solitary pulsars, or
those in binary systems where the orbital periods are longer than
about a day. For shorter-period binary systems, as noted by a number
of authors (see e.g.~Johnston \& Kulkarni \nocite{jk92} 1992), the
Doppler shifting of the pulse period results in a spreading of the
total signal power over a number of frequency bins in the Fourier
domain. Thus, a narrow harmonic becomes smeared over several spectral bins.

\begin{figure}[hbt]
\setlength{\unitlength}{1in}
\begin{picture}(0,2.7)
\put(-0.3,3.1){\includegraphics{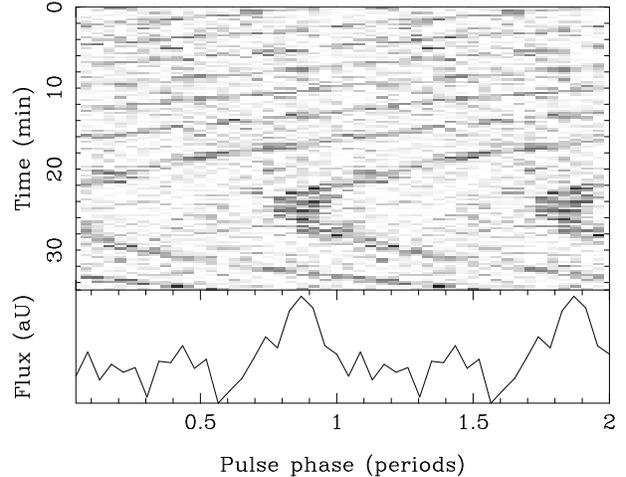}}
\end{picture}
\caption[]{
A search-mode observation of the eclipsing binary pulsar B1744--24A in
Terzan 5. Top panel: Data folded modulo the constant period reported by the 
search code (11.56 ms). The changing pulse period is clearly seen as a drift
in phase from one sub-integration to the next.  Lower panel: the
integrated pulse profile whose signal-to-noise ratio is significantly
degraded by the orbital motion during the integration.
}
\label{fig:ter5} 
\end{figure}

As an example of this effect, as seen in the time domain,
Fig.~\ref{fig:ter5} shows a 35-min search mode observation of PSR
B1744--24A; the 11.56 ms eclipsing binary pulsar in the globular
cluster Terzan 5 \nocite{lmd+90} (Lyne et al.~1990).  Given the short
orbital period of this system (1.8 hr), the observation covers about
one third of the orbit! Although the search code nominally detects the
pulsar with a signal-to-noise ratio of 9.5 for this observation, the
Doppler shifting of the pulse period seen in the individual
sub-integrations clearly results in a significant reduction in sensitivity.

The analysis reported in this paper makes no attempt to recover the
loss in sensitivity due to this effect. To date, the only pulsar
searches where this issue is tackled has been in searches for binary
pulsars in globular clusters \nocite{agk+90,clf+00} (e.g.~Anderson et
al.~1990; Camilo et al.~2000b). These searches applied a technique
whereby the time series is compensated for first-order Doppler
accelerations.  Although these searches have been very successful they
add significantly to the computational effort required to reduce the
data, and have therefore only been applied to globular clusters where
the DM is known a-priori from observations of solitary pulsars. For
our data, where the DM is a-priori unknown, we are presently
developing computationally-efficient algorithms which will permit us
to greatly improve the sensitivity to binary pulsars by re-analysing
these data in future.
We note that the present analysis results in significantly reduced
sensitivity to binary pulsars with orbital periods less than one day.

We conclude this discussion with some remarks on the search
sensitivity to very long-period ($P> 5$ s) pulsars. The existence of
radio pulsars with such periods are of great relevance to theories of
pulsar emission, many of which predict that the emission ceases when
the period crosses a critical value \nocite{cr93a} (see e.g.~Chen \&
Ruderman 1993). Young et al.~(1999) \nocite{ymj99}
have recently demonstrated that the period of PSR J2144$-$3933,
originally discovered in the Parkes Southern Sky Survey, is 8.5 s ---
three times that previously thought.  This is presently the longest
period for a radio pulsar. Young et al.~make the valid point that such
pulsars could be very numerous in the Galaxy since they have very
narrow emission beams and therefore radiate to only a small fraction
of the celestial sphere. An additional factor here is that the number
of pulses emitted by e.g.~a 10-s pulsar during typical pulsar survey
integration times is $\lapp 30$. If the pulsar undergoes
significant periods in the null state, as might be expected
\nocite{rit76} (Ritchings 1976), it will be harder to detect in
an FFT-based search \nocite{nic99} (Nice 1999).

One way to tackle this problem is to employ longer integration times,
such as we do here. The FFT-based periodicity search we use is,
however, not an ideal means to find long period signals since the
sensitivity is degraded by a strong ``red noise'' component in the
amplitude spectrum. The noise itself is a result of DC-level
fluctuations (e.g.~in the receiver) during an observation. In the
above analysis of the survey data, we minimised the effects of this
red noise component by subtracting a baseline off the spectrum before
normalising it. However, because of the rapid increase of the red
noise below about 0.1-Hz, we chose to ignore all spectral signals with
frequencies below this value. Whilst this is common practice in pulsar
search codes, it obviously reduces our sensitivity to $P>10$ s
pulsars! In recognition of this selection effect, we are currently
re-analysing our data using a so-called ``fast folding'' algorithm
\nocite{sta69} (e.g.~Staelin 1969) to search for periodic signals in
the period range 3--20 s. The results of this analysis, and a detailed
discussion of the algorithm, will be presented elsewhere (M\"uller et
al.~in preparation).

\section{Survey results}
\label{sec:res}

A total of seven pulsars were detected during the course of the
survey, four of which were previously unknown. Follow-up observations
carried out to confirm the existence of each of the new pulsars were
used to check that the true period had been correctly identified by
the search code. The basic properties and detection statistics of all
seven pulsars are summarised in Table \ref{tab:allpsrs}. Flux values
for the previously known pulsars are taken from \nocite{lylg95}
Lorimer et al.~(1995). Flux values for the newly discovered pulsars
are averages of a number of independent measurements based on the
timing measurements described in Sec.~\ref{sec:fup} and have
fractional uncertainties of about 30\% in each case.  The relative
positions of all these pulsars are shown on our sensitivity curve in
Fig.~\ref{fig:smin}.  

\begin{table}[hbt]
\begin{center}
\begin{tabular}{lllll}
\hline
PSR & $P$ & DM & $S_{1400}$ & S/N \\
    & (ms) & (cm$^{-3}$ pc) & (mJy) & \\
\hline
\multicolumn{5}{c}{Previously known pulsars} \\
\hline
B1842--02 & 507.7 & 429 & 1.0 & 21,7.0 \\
B1839--04 & 1840  & 196 & 8.5 & 34,41,85 \\
B1841--04 & 991.0 & 124 & 1.3 & 34 \\
\hline
\multicolumn{5}{c}{Newly discovered pulsars} \\
\hline
\EPSRD    & 204.1 & 170 & 0.9 & 21,32 \\
\EPSRA    & 526.7 & 167 & 0.5 & 13 \\
\EPSRB    & 525.1 & 908 & 0.6 & 7.5,10 \\
\EPSRC    & 207.7 & 500 & 0.6 & 8.0 \\
\hline
\end{tabular}
\end{center}
\caption[]
{
Basic parameters and search signal-to-noise ratios (S/N) for the seven
pulsars detected. Multiple S/N entries correspond to
detections in neighbouring grid positions.
}
\label{tab:allpsrs}
\end{table}

The astute reader will, by now, have noticed a striking similarity
between the periods of PSRs \EPSRA\, and \EPSRB\, and, to a lesser
extent, PSRs \EPSRC\, and \EPSRD. This unexpected result initially
gave us some cause for concern as to whether the signals we had
detected were indeed pulsars! However, having thoroughly investigated
each new pulsar, we are now confident that this is nothing more than a
bizarre coincidence. A number of independent facts confirm
this. Firstly, all the new pulsars are separated by a significant
number of telescope pointings on the sky. Secondly, the periods are
detected {\it only} at the nominal position of each pulsar, and
therefore cannot be put down to terrestrial interference. Furthermore,
all the dispersion measures are significantly different.
Finally, our timing measurements show that each pulsar
has a distinct set of spin-down parameters.

We note in passing that this survey places an upper limit to the
pulsed radio emission from the 6.97-s anomalous X-ray pulsar
J1845.0$-$0300 discovered by Torii et al.~(1998) that
lies in the search region. No radio pulsations were seen at the grid
position closest to this pulsar, setting a 1400-MHz pulsed flux limit
of $\sim 0.3 (\delta/4)^{1/2}$ mJy, where $\delta$ is the pulse duty cycle
in percent. This limit assumes (possibly incorrectly) that the effects
of interstellar scattering are negligible along this line of sight at
this observing frequency.  Deeper radio searches for this object, and
also for the 11.8-s pulsar in Kes~73 (Vasisht \& Gotthelf 1997), \nocite{vg97} 
should be carried out in future at different observing frequencies.

\section{Follow-up observations}
\label{sec:fup}

\begin{figure}[hbt]
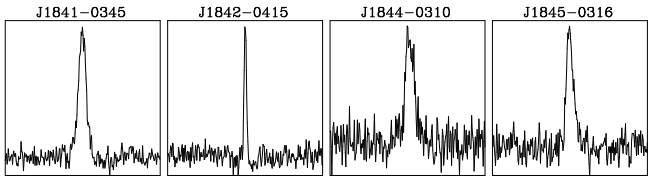

\setlength{\unitlength}{1in}
\begin{picture}(0,1)
\put(-0.10,1.1){\includegraphics{1830.f4a}}
\put(+0.75,1.1){\includegraphics{1830.f4b}}
\put(+1.60,1.1){\includegraphics{1830.f4c}}
\put(+2.45,1.1){\includegraphics{1830.f4d}}
\end{picture}
\caption[]
{
Integrated 1400-MHz pulse profiles for the four newly discovered
pulsars. The ordinate axis shows intensity, whilst the abscissa
shows rotational phase. Each profile displays 360 degrees of rotational phase.
These profiles are freely available in digital form at the European
Pulsar Network data archive (http://www.mpifr-bonn.mpg.de/div/pulsar/data).
}
\label{fig:profs}
\end{figure}

In order to obtain more detailed spin and astrometric parameters of
the newly-discovered pulsars, following confirmation, each was
included in our monthly $\lambda$ 21-cm timing observations of
millisecond pulsars using the Effelsberg-Berkeley-Pulsar-Processor. 
Full details of the observing procedures are described by
\nocite{kll+99} Kramer et al.~(1999). In brief, during each observing
session, a pulse time-of-arrival (TOA) measurement is obtained for
each pulsar by cross-correlating the observed pulse profile with a
high signal-to-noise ``template'' profile constructed from the
addition of many observations. The template profiles obtained in this
way are presented in Fig.~\ref{fig:profs}.

\begin{figure}[hbt]
\setlength{\unitlength}{1in}
\begin{picture}(0,2.6)
\put(-0.2,2.8){\includegraphics{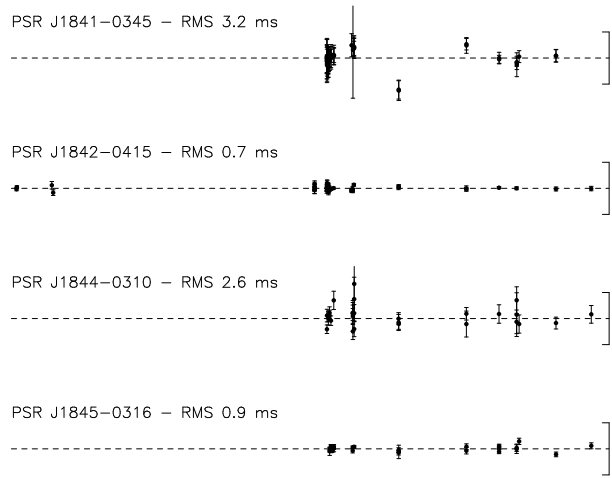}}
\end{picture}
\caption[]
{
Timing model residuals for the four newly discovered pulsars. The
timing baseline (dotted line) spans 440 days (MJD 51025-51465) in all
cases to accommodate early points gathered for PSR \EPSRA. Vertical
bars at the end of each plot denote $+/-$ 10 ms.
}
\label{fig:residuals}
\end{figure}

For each pulsar, the TOAs obtained from all the sessions were referred
to the equivalent time at the solar system barycentre and fitted in a
bootstrap fashion to a simple spin-down model using the {\sc tempo}
software package\footnote{Available from http://pulsar.princeton.edu/tempo}.
In Fig.~\ref{fig:residuals}, we present the resulting model-observed TOA
residuals from this analysis.

\begin{table*}[hbt]
\begin{center}
\begin{tabular}{lllll}
\hline
PSR              &      {\EPSRD}     &     {\EPSRA}     &{\EPSRB}&{\EPSRC} \\
\hline
R. A. (J2000)    &  $18^{\rm h} 41^{\rm m} 37.2(2)^{\rm s}$  &
                    $18^{\rm h} 42^{\rm m} 11.31(2)^{\rm s}$  & 
		    $18^{\rm h} 44^{\rm m} 45.7(2)^{\rm s}$  & 
		    $18^{\rm h} 45^{\rm m} 52.9(1)^{\rm s}$  \\
Decl. (J2000)    &  $-03^{\circ} 45^{'} 01(50)^{''}$        &
                    $-04^{\circ} 15^{'} 38.2(5)^{''}$        &
                    $-03^{\circ} 10^{'} 53(32)^{''}$        &
                    $-03^{\circ} 16^{'} 05(12)^{''}$        \\
$l$ (deg)        & ~~28.5  & ~~28.1   & ~~29.3  & ~~29.4\\
$b$ (deg)        & +0.47 &  +0.11 & +0.03  & $-$0.26\\
Period (s)     &  0.2040665324(4) &
                    0.52668125777(7) & 
                    0.5250489275(4) & 
                    0.2076357201(1) \\
Epoch  (MJD)     &  51330  &    51070  & 51350 & 51350  \\
$\dot{P}$ ($10^{-15}$ s/s)& 58.9(3)  &    21.94(1)  & 10.1(4) & 8.86(6)  \\
DM (cm$^{-3}$\,pc) &  170(5) &    167(1) & 910(30) & 500(5)  \\
Data Span (MJD)  & 51257-51451 & 51028-51451 & 51257-51451& 51259-51451\\
Number of TOAs   &   45      &      42   &    29      &   24    \\
RMS residual (ms)&   3.2    &       0.7  &    2.6     &   0.9     \\
\hline
$\tau_c$ (kyr)   &    55     &       380   &  380   &  371   \\ 
$B$ ($10^{12}$ G)&    3.5    &        3.4  &   2.3  &   1.4   \\
$D$ (kpc)        &    3.7   &  3.6  & 9.4 & 6.5 \\
$L_{1400}$ (mJy kpc$^2$) & 12.5 & 6.0 & 56.6 & 24.4 \\
\hline
\end{tabular}
\end{center}
\caption[]
{
Parameters for the four newly discovered pulsars.
Values in parentheses are uncertainties in the least significant digit.
}
\label{tab:newpsrs}
\end{table*}

The phase-coherent timing solutions we obtain for each pulsar indicate
that they are all solitary objects. The fitted parameters are
summarised in Table \ref{tab:newpsrs}. A sub-arcsecond position has
been determined for PSR \EPSRA, where the baseline of timing
observations already spans over a year. The remaining pulsars have
timing baselines spanning just over 6 months. This is however,
sufficient to decouple the covariant effects of position error and
spin down and, as a result, we have determined accurate period
derivatives for each pulsar. Table \ref{tab:newpsrs} also lists the
characteristic ages ($\tau_c$) and surface magnetic field strengths
($B$) inferred from these measured period and period derivatives (see
e.g.~Manchester \& Taylor 1977 for definitions of these \nocite{mt77}
parameters). In addition, we also list the distance ($D$) to each
pulsar inferred from its DM, Galactic coordinates and the Taylor \&
Cordes (1993) \nocite{tc93} electron density model, as well as the
1400-MHz luminosities inferred from these distances and the observed
flux densities as $S_{1400} D^2$.

It is significant that five of the seven pulsars detected in this
survey (including all the newly-discovered pulsars) have
characteristic ages below 0.5 Myr --- over an order of magnitude
younger than the median age of the normal pulsars detected by the
Parkes 70-cm Southern Sky survey (Manchester et al.~1996; Lyne et
al.~1998). This result should not be surprising when it is realised
that we have preferentially selected a sample of objects located close to 
their birth sites along the Galactic plane (Clifton et al.~1992; 
Johnston et al.~1992a).

By far the youngest of the new discoveries is PSR \EPSRD, which has a
characteristic age of only 55 kyr. Since this is within the mean
lifetime of supernova remnants \nocite{fgw94} (60 kyr --- Frail et al. 1994),
we checked the position of this and the other newly
discovered pulsars with the most recent catalogue of supernova
remnants \nocite{gre98} (Green 1998) for spatial coincidences. No
supernova remnants in the catalogue lie within 0.3 degrees of \EPSRD,
or indeed any of the other new pulsars.

In their study of pulsar-supernova remnant associations
Frail et al.~(1994) undertook a programme of deep radio
imaging to search for previously undetected supernova remnants around
several of the young pulsars from the Johnston et al.~(1992a) survey.
Using the accurate positions we obtained from the timing analysis, we
examined the NRAO VLA Sky Survey (NVSS; Condon et al.~1998)
\begin{figure}[hbt]
\setlength{\unitlength}{1in}
\begin{picture}(0,3.7)
\put(0.0,0.0){\includegraphics{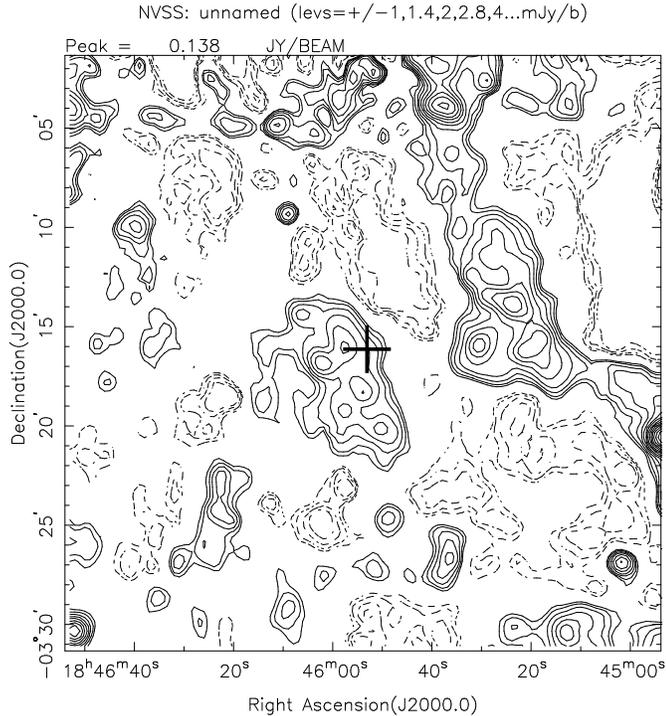}}
\end{picture}
\caption[]
{
NVSS image of the field surrounding PSR \EPSRC. The pulsar position
is marked by the $+$ sign at the rim of an extended radio source.
}
\label{fig:nvss}
\end{figure}
\nocite{ccg+98} images of the fields surrounding each pulsar for
evidence of diffuse $\lambda$ 20-cm emission which could be attributed
to uncatalogued supernova remnants. The only pulsar for which any
diffuse emission is evident in the NVSS survey (down to the 1-mJy
sensitivity limit) is \EPSRC, shown in Fig.~\ref{fig:nvss}.  It is
presently not at all obvious whether this emission is attributable to
the supernova remnant associated with this pulsar simply because there
is such a high density of similar radio sources in this region of the
sky. As a result, the by-chance probability of finding unrelated
diffuse radio emission, particularly in deeper images of this region,
will be rather high, making it difficult to unambiguously identify any
associated supernova remnants without additional information 
(e.g.~independent distance estimates to the pulsar and the candidate
remnant).

\section{Conclusions}
\label{sec:conc}

The discovery of four sub-mJy pulsars in the limited pilot search
observations reported here clearly demonstrate the potential for
future pulsar surveys with the Effelsberg radio telescope. As
mentioned earlier, the main aim of this survey was to test the
feasibility of finding pulsars with a new wide-band search system
currently under development. This new system employs narrower channel
bandwidths and has much faster sampling rates than presently
available; it will therefore have significantly improved sensitivity
to short-period, highly dispersed pulsars.

Now that the Parkes multibeam survey is extending its coverage out to
$l=50^{\circ}$ \nocite{lcm+00} (Lyne et al.~2000) there is little to
be gained in using the new system at Effelsberg to initiate a
large-scale $\lambda$ 21-cm search of the Galactic plane.  A targeted
$\lambda$ 21-cm search of globular clusters, however, is a worthy
scientific goal since deep (several hour) integrations would achieve a
substantially improved sensitivity over previous searches
\nocite{bl96} (see e.g.~Biggs \& Lyne 1996). Such a search would be
particularly timely given the flurry of binary pulsar
discoveries in a recent $\lambda$ 21-cm search of 47~Tucanae
\nocite{clf+00} (Camilo et al.~2000b).

Another excellent use of the new system would be an $\lambda$ 11-cm
search for heavily scattered pulsars close to the plane. Such a search
would open up an entirely new area of parameter space in Galactic
plane searches since it is known that many pulsars discovered at
$\lambda$ 21-cm are still strongly affected by interstellar
scattering. The strong inverse dependence of scattering on observing
frequency means that the effects of scattering on an $\lambda$ 11-cm
search would be an order of magnitude smaller than at $\lambda$ 21 cm.
In the vicinity of the Galactic centre, where scattering is expected
to be greatest \nocite{cl97} (Cordes \& Lazio 1997), the best prospects 
for finding pulsars still seem to be in searches carried out at 5 GHz 
($\lambda$ 6-cm), or even higher frequencies (see e.g.~Kramer et 
al.~1996; Kramer et al.~2000).

\acknowledgements 
We wish to thank Jiannis Seiradakis for help and encouragement during
the survey observations, and Oleg Doroshenko for assistance with the
timing observations. The constant support of the skilled and dedicated
operators at Effelsberg {\it viz:} Herrn. Koch, Marschner, Seidel,
Georgi, Schlich and Bartel was instrumental in our quest to finally
find pulsars with the Effelsberg radio telescope. DRL would also like
to thank Chris Salter for useful comments on an earlier version of the
manuscript, and Bryan Gaensler for helpful discussions.

\end{document}